%% file: main.tex
\documentclass[11pt,a4paper]{article}

\usepackage[
  left=3.2cm,
  right=3.2cm,
  top=3.0cm,
  bottom=3.0cm
]{geometry}

\usepackage[T1]{fontenc}
\usepackage[utf8]{inputenc} 
\usepackage{lmodern}        
\usepackage{microtype}      

\usepackage{setspace}
\usepackage{graphicx}
\usepackage[font=small,labelfont=bf]{caption}
\usepackage{titlesec}
\titleformat{\section}{\large\bfseries}{\thesection}{1em}{}
\titleformat{\subsection}{\normalsize\bfseries}{\thesubsection}{1em}{}

\usepackage{array}

\title{Better Balance in Informatics 2.0: \\
The First-Year Students}
\author{Ine Arvola \and Rakel Håndlykken \and Elisavet Kozyri}
\date{Department of Computer Science \\ 
UiT The Arctic University of Norway}

\usepackage[numbers]{natbib}
\begin{document}

\maketitle

\begin{abstract}
Diversity among computer scientists and technologists is necessary for the sustainable development of society through technological innovation. At UiT The Arctic University of Norway, only 13\% of computer science students are women. Many find the learning curve in introductory computer science courses to be very steep, and thus, they drop out. Female students tend to be overrepresented in this group.
The goal of this project was to improve the gender balance among computer science students at UiT by focusing on female first-year students and ensuring that they do not drop out of the study programs in the first year of study. The project established a seminar series for strengthening the basic programming-technical skills that many first-year students lack, and exposing them to different aspects and career paths within the computer science subject beyond the focus area of the study program.
Results show positive developments, particularly related to the students' perceived introduction to basic technical topics. A comparison between 2024 and 2025 shows improvements in several of the areas addressed in the technical workshops, including use of file systems, terminals, debugging and the code development process.
However, effects on dropout and study experience require more long-term measures.
\end{abstract}

\section{Introduction}
\input{sections/introduction}

\section{Methodology}
\input{sections/Methodology/technical}

\input{sections/Methodology/cultural}

\input{sections/Methodology/promotion}

\section{Results of the Impact Survey}
\input{sections/survey2425_results}

\section{Dropout and Coursework Data}
\input{sections/course_results}
\section{Discussion of Impact Survey, Dropout, and Course Data}
\input{sections/analysis_of_results}

\section{Evaluation of BBI 2.0}
\input{sections/evaluation}

\section{Related Work}
\input{sections/related_work}

\section{Our Recommendations}
\input{sections/conclusion}

\bibliography{references}

\end{document}

%% file: sections/introduction.tex
Early experiences in Computer Science education play a critical role in shaping students’ confidence, motivation, and persistence in the field. Previous research has shown that students with limited prior exposure to computing, particularly females, are at higher risk of feeling academically behind and disengaging early in their studies. These challenges are often reinforced by steep learning curves in introductory courses and limited visibility of the breadth of Computer Science (CS) as a discipline.

\emph{Better Balance in Informatics 2.0} (BBI 2.0)\footnote{https://uit.no/project/bbi2},
which was funded by the Research Council of Norway,\footnote{Project 350465, BBI 2.0: Bedre Balanse i Informatikk - En mer inkluderende introduksjon til informatikk}
was developed to address these challenges and contribute to improved gender balance among CS students. The primary goal of the project was to reduce early dropout among first-year CS students, with particular attention to retaining female students during their first encounter with university-level programming. In addition, the project aimed to improve students’ perceived preparedness for introductory courses, strengthen their confidence in working with technical material, and increase satisfaction with their choice of study by exposing them to a broader range of topics and career paths within Computer Science. 

To support these goals, BBI 2.0 was implemented as a supplementary seminar series running alongside the first semester of the study program in the Fall of 2025. The initiative was motivated by the need to support female students and was primarily designed with their challenges in mind, particularly regarding gaps in prior technical experience and uncertainty about belonging in the field. At the same time, participation in the seminar series was open to the entire student cohort, and all first-year students were encouraged to attend.

The seminar series consisted of two complementary components: technical and cultural. The technical workshops focused on strengthening foundational skills that are often assumed but not explicitly taught in introductory courses, such as working with the file system, using the terminal, debugging code, and version control. The cultural seminars aimed to broaden students’ understanding of Computer Science by presenting diverse applications of the field, introducing role models from academia and industry, and highlighting potential career trajectories.

This report evaluates the BBI 2.0 seminar series using impact-survey data collected in 2024 and 2025, as well as responses from a follow-up evaluation survey distributed after the completion of the seminar series. In addition, participation records from the seminars and contextual exam results are included. The evaluation focuses on indicators related to the project’s goals, including students’ perceived preparedness, confidence, satisfaction with their study choice, and intentions to continue studying CS, as well as students’ experiences with and perceptions of the seminar series itself. The results are interpreted in relation to the broader structural context of the introductory courses and the overall objectives of the BBI 2.0 initiative.

%% file: sections/Methodology/technical.tex
This section describes how BBI 2.0 was designed and carried out. The project consisted of two main components: a technical seminar series and a cultural seminar series. Both were voluntary, they were held in the Fall semester of 2025 and aimed at supporting first-year students academically and socially.
In addition to describing the content and structure of the events, this section also outlines how the series was promoted and how participation was recorded.

\subsection{Technical Seminar Series}
The technical part of BBI 2.0 consisted of a series of voluntary workshops designed to support first-year students in developing foundational skills in CS. Each workshop focused on a specific topic that is implicitly required in introductory programming courses, but not always explicitly taught. The topics included file systems, terminal usage, code development practices, debugging, report writing, and version control using GitHub.

The motivation for choosing these topics came from discussions with lecturers and senior students, who pointed out recurring gaps in basic knowledge among first-year students. The presenters were mainly senior students who were active in the student environment and had previous experience as teaching assistants. There was also a deliberate effort to include female presenters, both for representation and visibility.

The workshops combined short theoretical introductions with hands-on exercises. The table below presents an overview of the workshops, including dates, themes, and brief descriptions.

\begin{table}[h!]
\centering
\begin{tabular}{p{2.5cm} p{4cm} p{7cm}}
\hline
\textbf{Date} & \textbf{Theme} & \textbf{Description} \\
\hline

20/08 & File System and Shell Environment & Introduction to how file systems are structured and how the terminal interacts with the computer. The workshop aimed to ensure that all students developed a basic understanding of their own laptop before beginning programming tasks. \\

26/08 & Code Design and Report Writing & The first part focused on translating problem-solving approaches into structured code, helping students identify what is important in a task and how to systematically work toward a solution. The second part introduced report writing, source usage, and source criticism in an AI-influenced academic context, clarifying expectations and responsible use of AI tools. \\

02/09 & Code Development Cycle and Debugging & An introduction to writing, compiling, running, and refining code using text editors and the terminal. The workshop also included practical training in identifying and resolving common programming errors, with the aim of building confidence in systematic debugging and problem-solving. \\

09/09 & GitHub and Version Control & Introduction to version control concepts and collaborative coding using Git and GitHub. Students practiced managing repositories, tracking changes, and working efficiently in team-based development workflows. \\

14/10 & PC components &  An introduction to how a computer is built and how its components work together. The workshop included hands-on activities where students disassembled laptops and desktops to identify key components, as well as learn basic maintenance, troubleshooting, and upgrading techniques. \\
\hline
\end{tabular}
\caption{Overview of the BBI 2.0 technical workshops}
\end{table}

%% file: sections/Methodology/cultural.tex
\subsection{Cultural Seminar Series}
The cultural seminar series aimed to strengthen students’ sense of belonging and broaden their understanding of CS as a field. These events consisted of invited talks by researchers and industry professionals, highlighting different applications of CS and a variety of career paths.

The selection of speakers and topics was based on discussions with faculty members and experiences from earlier BBI initiatives. Emphasis was placed on choosing relevant and inspiring themes, while also ensuring the visibility of diverse role models.

An open call was sent out within the department inviting suggestions for presenters. After compiling a long list of potential speakers, selections were made based on thematic relevance and with the aim of including contributors from both industry and academia, as well as from local, national, and international contexts. Most events were structured as presentations followed by discussion, with one session organized as a student panel debate.

Table~\ref{fig:Overview cultural events} provides an overview of the cultural events, including the date, theme, and a brief description of each session.

\begin{table}
\centering
\begin{tabular}{p{2.5cm} p{4cm} p{7cm}}
\hline
\textbf{Date} & \textbf{Theme} & \textbf{Description} \\
\hline

16/09 & Panel Debate: What Is It Really Like to Study Computer Science? & A panel with students from different years sharing honest experiences about academic workload, expectations, social environment, and coping with pressure. The session encouraged open discussion and audience participation. \\

26/09 & How to Secure Your Dream Job & A personal reflection on building a meaningful career without a clear long-term plan. The talk highlighted career flexibility in technology and showed how opportunities can emerge unexpectedly through experience, curiosity, and initiative. \\

02/10 & Artificial Intelligence for All & A lecture discussing diversity, inclusion, and ethical challenges in AI and software engineering. The presentation emphasized the risks of bias in AI systems and the importance of intersectionality and representation when developing future technologies. \\

07/10 & Brain Orchestra in Resting-State & A research-based lecture on computational neuroscience, presenting how large-scale neuronal networks can be analyzed using computer science methods. The talk demonstrated interdisciplinary applications of informatics in understanding brain dynamics and complex systems. \\

21/10 & From Nerd Niche to Foundation of Modern Society & A reflection on over 40 years of informatics development, showing how research and education at UiT have contributed to startups, search engine technology, sports analytics, AI, and medical diagnostics. The talk highlighted long-term career possibilities and societal impact. \\

28/10 & This Was Not in the Study Plan & A presentation about unexpected career paths in informatics, from starting a tech startup to later choosing a stable industry position. The talk emphasized that a CS degree opens diverse opportunities beyond traditional developer roles. \\

11/11 & From Academia to Startup & A journey from academic research to founding a technology startup, reflecting on the transition from theory to practice. The presentation explored what academia prepares students for, what it does not, and how entrepreneurial skills are developed outside the classroom. \\

25/11 & IT Leadership in Industry & A career story illustrating how technical expertise can evolve into leadership roles. The talk described experiences from software development in industry to strategic IT management within large organizations. \\

\hline
\end{tabular}
\caption{Overview of BBI 2.0 cultural seminar series}
\label{fig:Overview cultural events}
\end{table}

%% file: sections/Methodology/promotion.tex
\subsection{Promotion}
This section describes the strategies and channels used to reach students and promote the project’s activities. In the promotion strategy it was important to ensure early recruitment to the series. To introduce the project to first-year students, senior students were included in the outreach strategy. Promotion was carried out through the communication channels of Tromsøstudentenes Dataforening (TD), both the project’s own communication channels and the BBI website were advertised. The website served as the primary information hub for the project, publishing updates about lectures and workshops.

The main promotional strategy focused on recruiting first-year students early in the semester, as lecture attendance was expected to decline rapidly as the semester progressed. The project team participated in orientation meetings for all relevant study programs included in the course. During these meetings, students were asked to register their participation for the entire series. This registration data was used to promote workshops and events throughout the semester.

\subsubsection{Promoting Technical Series}
The promotion strategy for technical workshops focused on direct interaction with students and recruitment within their academic environments. This included promotion during lectures and study sessions with senior teaching assistants. The technical developers who led the workshops also contributed to the promotion efforts through their roles as senior students.
 Before each workshop, two promotional emails were sent to students registered for the series. These emails provided information about the workshop content and background information on the technical developers. Food was served at all technical workshops to further encourage student participation.

\subsubsection{Promoting Cultural Series}
The promotion strategy for cultural events involved sending two formal invitations to registered participants, publishing the invitations on the series’ private communication channels, and posting announcements on the university’s learning platform. The invitations included information about the presenter, an abstract of the talk, and encouragement for students to attend and actively engage by asking questions.
To increase attendance at events featuring international presenters and events that were during inconvenient times of the day, food was served and the events were opened to participants from the entire faculty.

\subsection{Collecting results}
To evaluate the extent to which the BBI 2.0 series achieved its intended goals, an impact survey was conducted in the Fall semester of 2024 and 2025. The survey was distributed late in the students’ first semester in CS and included questions designed to measure the effects of both the technical workshops and the cultural initiatives. For the 2024 cohort, students had not been exposed to the BBI 2.0 series. Therefore, the survey questions were formulated to be understandable without prior knowledge of the initiative, while still capturing whether students experienced outcomes aligned with its intended impact. This allowed for a baseline comparison with the 2025 cohort, which did participate in the BBI 2.0 activities. The 2024 and 2025 cohorts received the same survey.

In addition, an evaluation survey was distributed to the 2025 students who participated in the BBI 2.0 series, with the aim to provide additional context to the impact survey results.

Both surveys were distributed via email and actively promoted during lectures as well as at all BBI events. The results from both surveys will be presented and discussed in later sections of this report.

%% file: sections/survey2425_results.tex
This section presents the results from the BBI 2.0 impact surveys conducted in 2024 and 2025. The results are grouped to highlight patterns related to student satisfaction, perceived preparedness, confidence, retention intentions, and perceived exposure to inspiring topics and role models. Participation data from the seminar series is reported where relevant to contextualize the findings.

The results are presented descriptively. Interpretation and discussion of potential explanations and limitations are deferred to Section~5.

The impact survey was distributed in late November in both 2024 and 2025, at the end of the semester but prior to the exam period. This timing was chosen to avoid overlap with course evaluations and to ensure that students were still present on campus. It also allowed us to effectively promote the survey by attending lectures and engaging with students in common areas during this period. As shown in Table~\ref{tab:survey_participation}, participation in the survey decreased from 2024 to 2025. 

\begin{table}[htbp]
    \centering
    \begin{tabular}{l c}
    \hline
        Year & Impact survey participation \\
    \hline
        2024 & 29 \\
        2025 & 20 \\
    \hline
    \end{tabular}
    \caption{Impact survey participation by year}
    \label{tab:survey_participation}
\end{table}

The results from the survey are presented using bar charts to give an overview of how students responded to each question. Each plot corresponds to one survey question, and the horizontal axis shows the different response categories (e.g., “to a small extent”, “to a large extent”). The vertical axis represents the proportion or number of students selecting each category.  Responses are separated by gender to highlight potential differences between groups: "F/NB" corresponds to 
students that identify themselves as female or non-binary
and "M" corresponds to those identify themselves as male.

The plots are intended to provide a descriptive overview rather than precise measurements. They should be read together with the accompanying text, which explains the intention behind each question and highlights the most relevant findings. 

\subsection{Technical Series Results}
The following questions assess whether students felt they received adequate introductions to specific technical topics. These topics were directly addressed through the BBI technical workshops, making the questions a direct measure of the workshops’ perceived impact.

For each workshop, participation numbers are reported to provide context for the observed results. Participation decreased over the semester as the workshops were held later in the term, which is taken into account when interpreting the results.

\begin{table}[htbp]
    \centering
    \begin{tabular}{l c}
    \hline
        Workshop & Participation \\
    \hline
        File system/Shell & 54 \\
        Code design/report & 51 \\
        Code dev./debugging & 39 \\
        Github & 29 \\
        PC components & 16 \\
    \hline
    \end{tabular}
    \caption{Participation technical workshops}
    \label{tab:placeholder}
\end{table}

\begin{figure}[!hbt]
    \centering
    \includegraphics[width=0.75\linewidth]{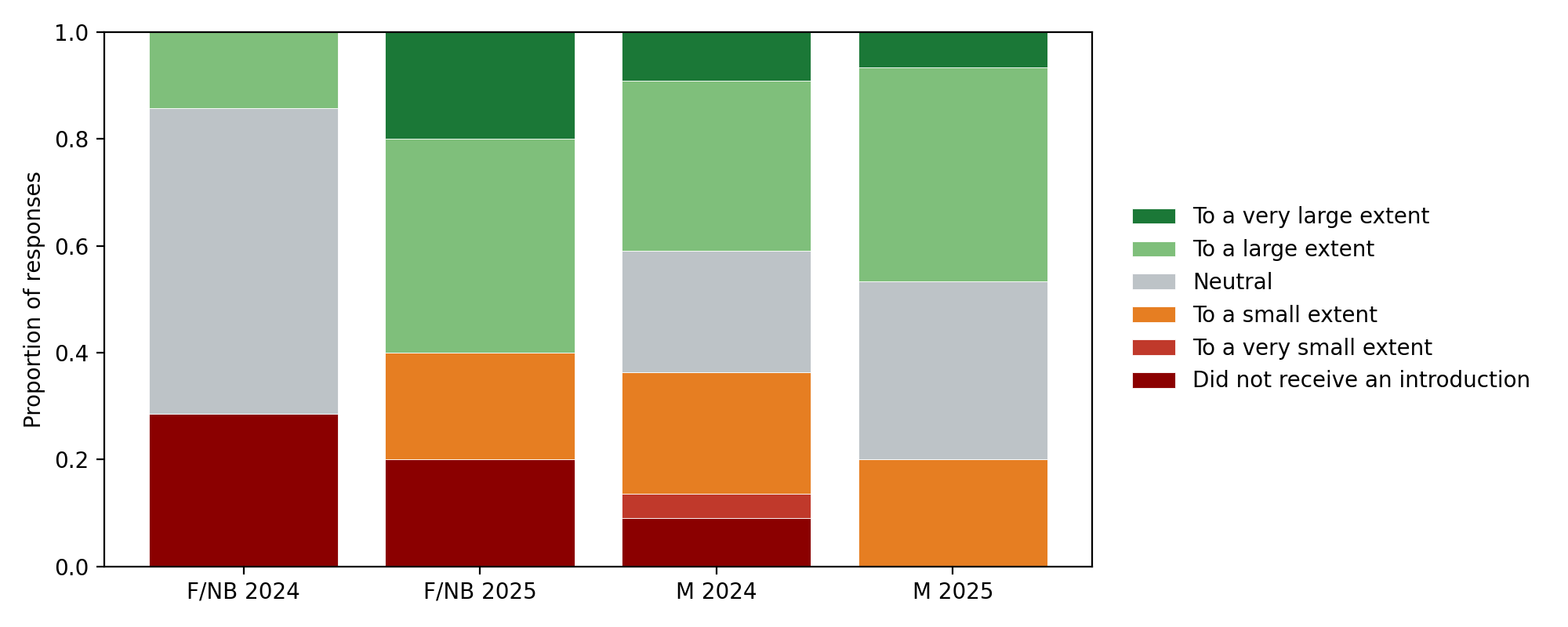}
    \caption{Introduction to File System Usage}
    \label{FILE SYSTEM}
\end{figure} 

\paragraph{Q1: To what extent do you feel that you have received a good introduction to using the file system?}
This question measures students’ perceived preparedness in file system usage, a foundational technical skill in the program. From 2024 to 2025, Figure \ref{FILE SYSTEM} shows that responses shift toward more positive categories for both genders. Female students show a clear movement away from neutral and negative responses, and male students report fewer low assessments. The gender difference visible in 2024 appears reduced in 2025. Overall, perceived introduction to file system usage improves.

\begin{figure}[!hbt]
    \centering
    \includegraphics[width=0.75\linewidth]{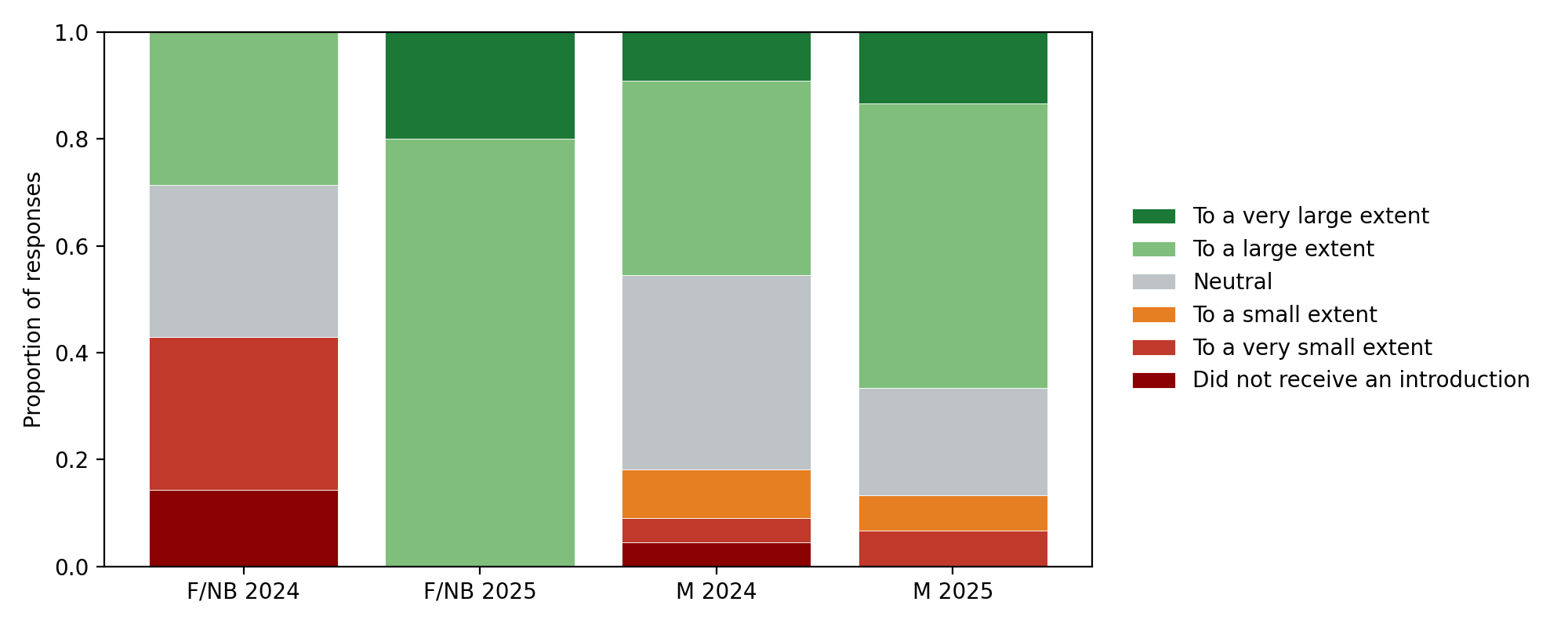}
    \caption{Introduction to Terminal Usage}
    \label{fig:term}
\end{figure}
\paragraph{Q2: To what extent do you feel that you have received a good introduction to using the terminal?}
This question assesses perceived introduction to terminal usage, another core technical component early in the semester. According to Figure \ref{fig:term}, compared to 2024, 2025 shows a substantial shift toward more positive responses. In 2024, a considerable share of female students reported limited introduction. In 2025, no female students report receiving no introduction, and most responses are in the higher categories. Male students also show improvement. The overall perception of terminal preparedness strengthens from 2024 to 2025.

\begin{figure}[!hbt]
    \centering

    \includegraphics[width=0.75\linewidth]{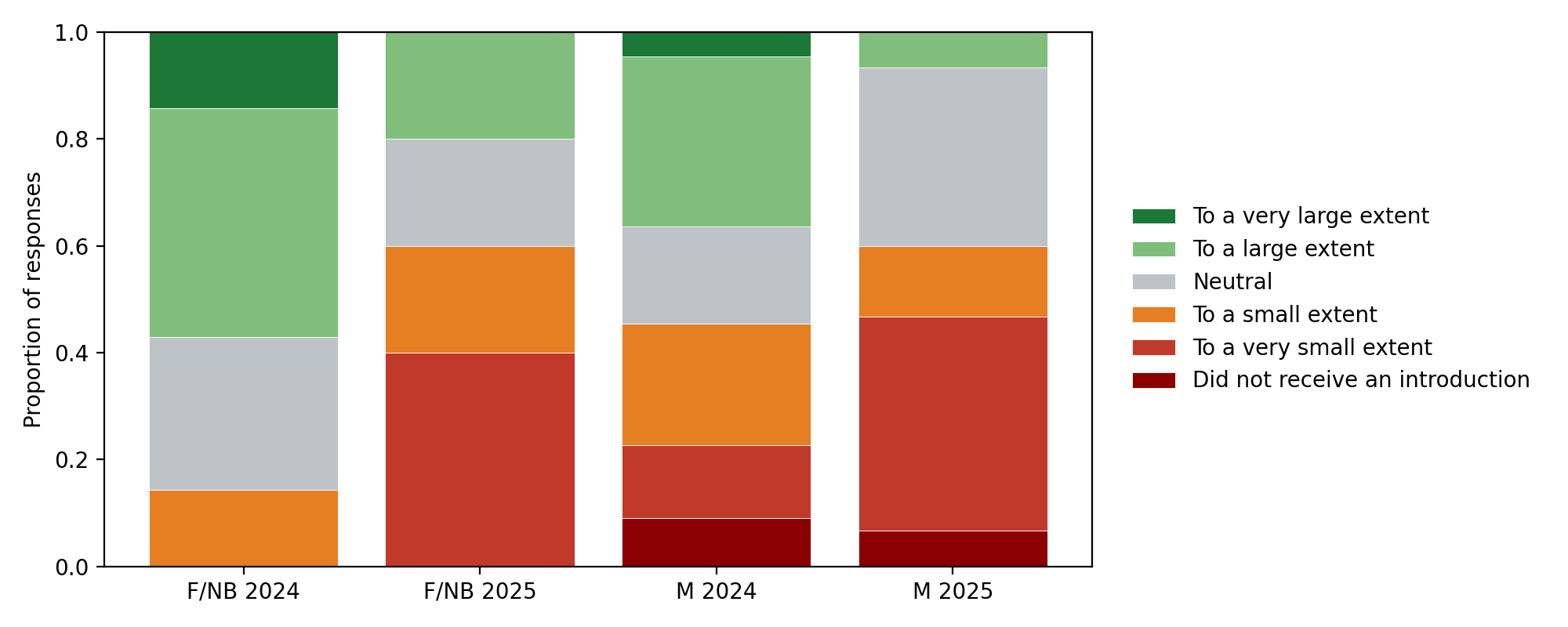}
    \caption{Introduction to Writing Reports}
    \label{report}
\end{figure}
\paragraph{Q3: To what extent do you feel that you have received a good introduction to writing reports?}
This question evaluates students’ perceived introduction to academic writing and report production, which are essential for coursework and assessment. Unlike the other technical topics, report writing does not show a clear positive development from 2024 to 2025, according to Figure \ref{report}. Male students’ responses remain concentrated in the lower and neutral categories. Among female students, responses shift slightly toward more negative categories, although a substantial share remain neutral or positive. Overall, perceived preparedness in report writing does not improve between the two years, and this will be discussed futher in the discussion.

\begin{figure}[!hbt]
    \centering
    \includegraphics[width=0.75\linewidth]{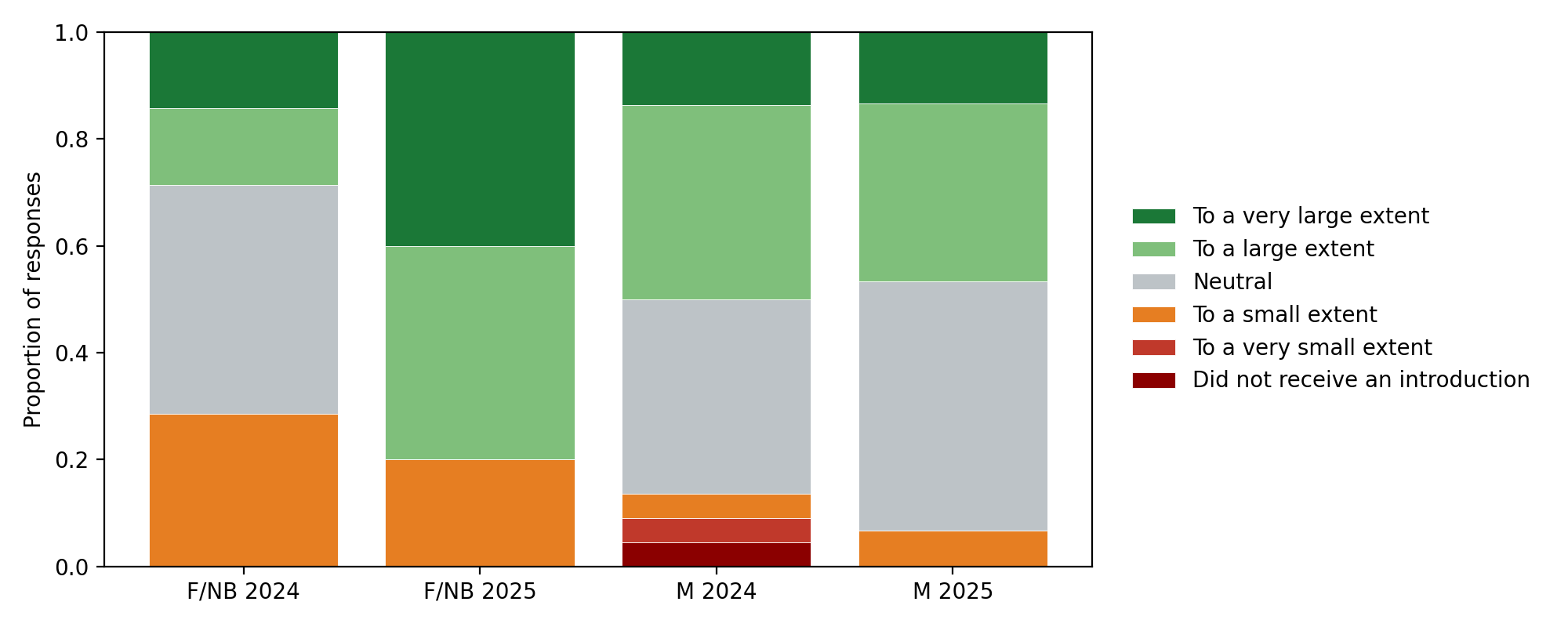}
    \caption{Introduction to the Code Development Process}
    \label{fig:codedev}
\end{figure}

\paragraph{Q4: To what extent do you feel that you have received a good introduction to the code development cycle?}This question measures whether students feel they understand the overall process of developing, testing, and refining code. Based on Figure \ref{fig:codedev}, from 2024 to 2025, female students show a shift toward higher categories, indicating improved perceived understanding of the development process. Male students remain relatively stable and positive across both years. The difference between genders visible in 2024 appears smaller in 2025. Overall, perceived introduction to the development process improves.

\begin{figure}[!hbt]
    \centering
    \includegraphics[width=0.75\linewidth]{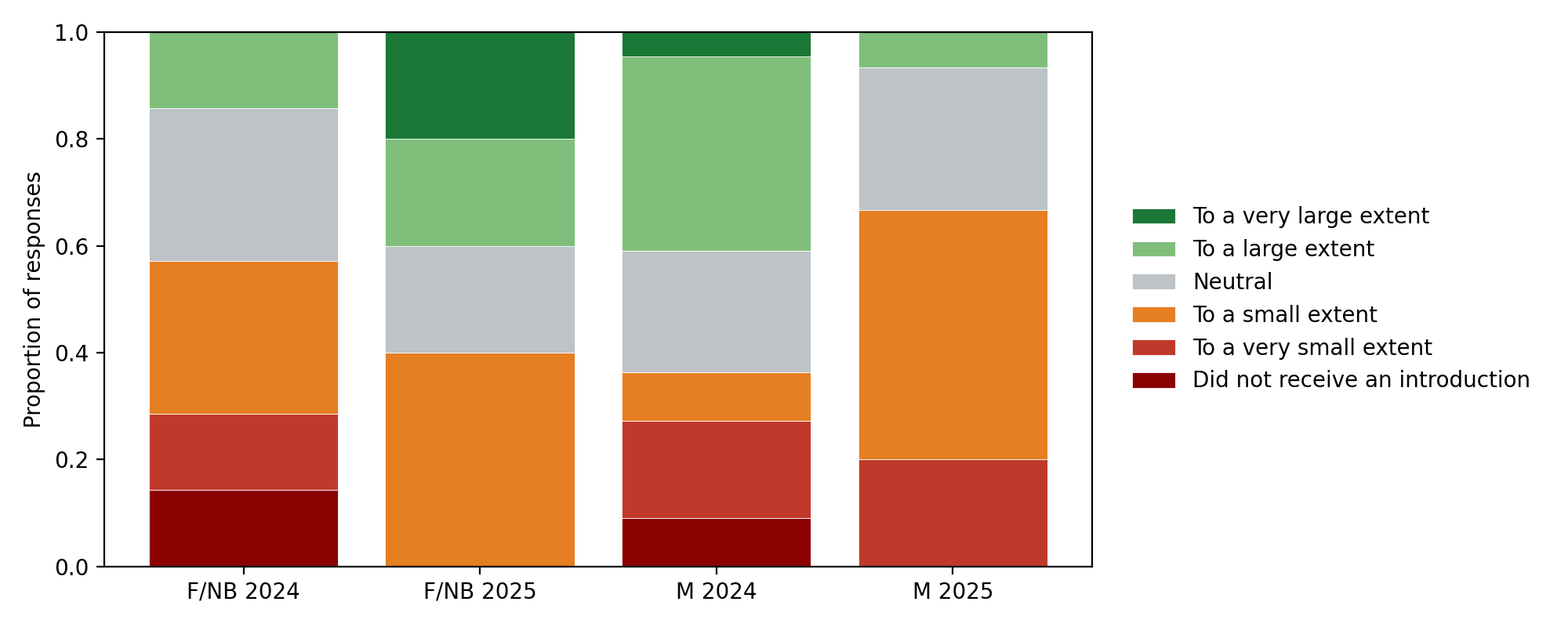}
    \caption{Introduction to Debugging}
    \label{fig:debug}
\end{figure}
\paragraph{Q5: To what extent do you feel that you have received a good introduction to debugging?}
This question assesses students’ perceived introduction to debugging, a skill often identified as challenging for beginners. Figure \ref{fig:debug} shows that, compared to 2024, responses in 2025 show improvement among female students. No students report receiving no introduction in 2025, and responses shift toward neutral and positive categories. Male students does not report any improvement. Overall, perceived preparedness in debugging increases for female students and decreased for male student from 2024 to 2025.

\begin{figure}[!hbt]
    \centering
    \includegraphics[width=0.75\linewidth]{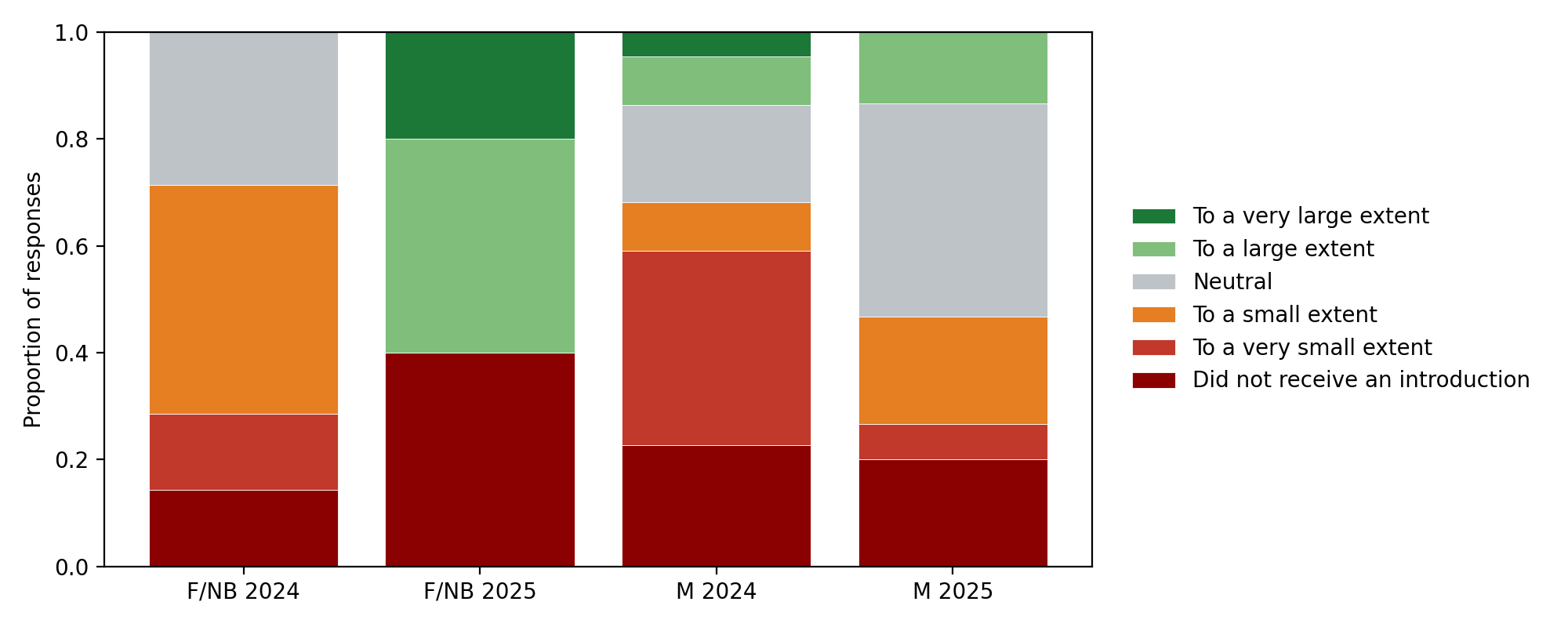}
    \caption{Introduction to GitHub}
    \label{fig:github}
\end{figure}
\paragraph{Q6: To what extent do you feel that you have received a good introduction to GitHub?}
This question measures students’ perceived introduction to GitHub, a tool that is widely used but often unfamiliar to first-year students. According to Figure \ref{fig:github}, in 2024, responses were largely concentrated in the lower and neutral categories across both genders. In 2025, among students who did attended the workshop, responses shift toward more positive categories, especially among female students. However, participation was lowest for this workshop, and many students reported that they did not receive an introduction, which means they did not attend. This again limits the impact.

\subsubsection*{Conclusion about the impact of the technical series}
Across the technical measures, the comparison between 2024 and 2025 shows a generally positive development, especially among female students. Improvements are most clearly visible in terminal usage, file system usage, debugging, and the code development cycle. In several of these areas, responses shift toward higher categories for both genders, but most significantly among female students, and the gender differences observed in 2024 appear reduced in 2025.

Report writing stands out as the only technical area that does not show a comparable improvement. In this case, responses remain stable or slightly more negative in 2025.

\subsection{Cultural Series Results}
The following questions assess whether students experienced a sense of inclusion, belonging, and representation within the study environment. These aspects were directly addressed through the cultural seminars.

Participation numbers for each cultural event are included to contextualize the results. As the semester progressed, attendance declined for seminars held later in the term, and this trend is considered when interpreting the findings.

\begin{table}[htbp]
    \centering
    \begin{tabular}{l c}
    \hline
        Seminar theme & Participation \\
    \hline
        Panel Debate & 6 \\
        How to Secure Your Dream Job & 6 \\
        Artificial Intelligence for All & 17 \\
        Brain Orchestra in Resting-State & 30 \\
        From Nerd Niche to Foundation of Modern Society & 20 \\
        This Was Not in the Study Plan & 10 \\
        From Academia to Startup & 16 \\
        IT Leadership in Industry & 1 \\
    \hline
    \end{tabular}
    \caption{Participation in cultural seminars and panel debate}
    \label{tab:seminar_participation}
\end{table}

\begin{figure}[!hbt]
    \centering
    \includegraphics[width=0.75\linewidth]{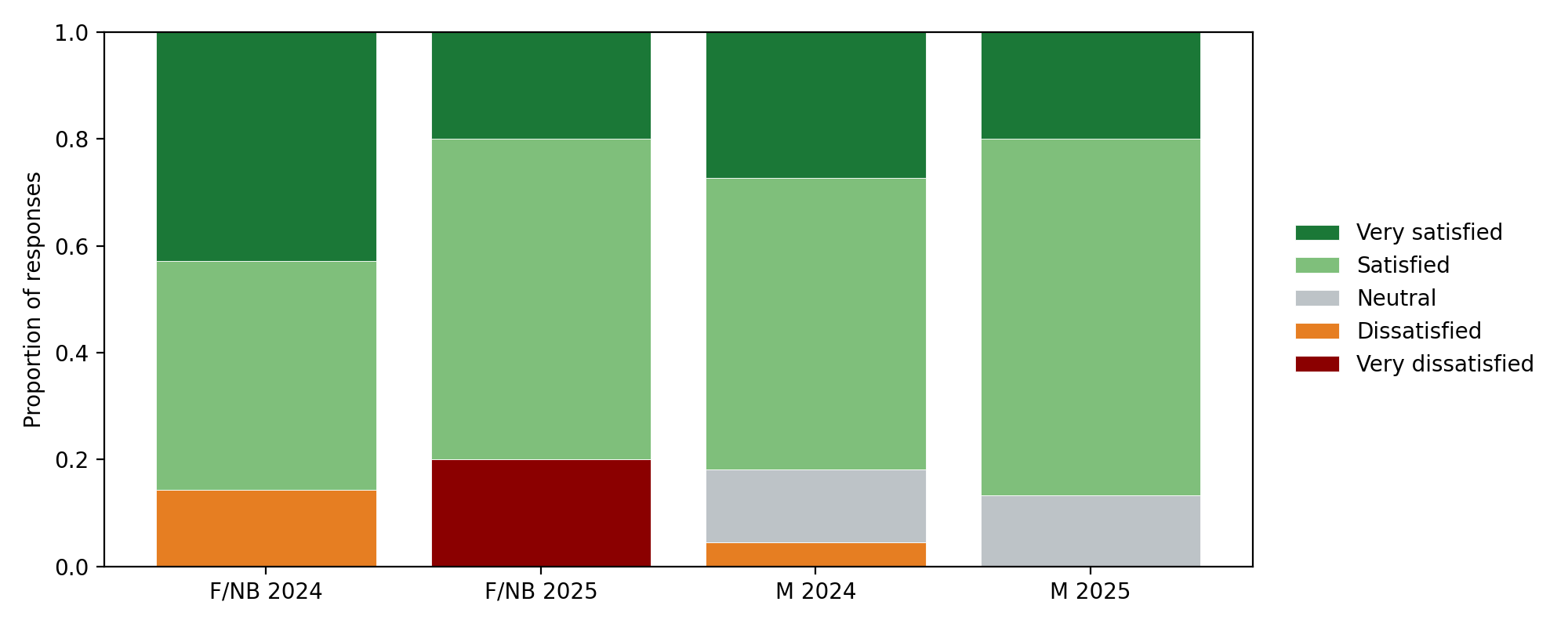}
    \caption{Satisfied with decision to study computer science}
    \label{cssat}
\end{figure}
\paragraph{Q7: To what extent do you feel satisfied with your decision to study computer science?}
 This question was included to measure students’ overall satisfaction with their choice of study, as dissatisfaction early in the program is a known risk factor for dropout. 
 According to Figure \ref{cssat}, from 2024 to 2025, satisfaction increases among male students, while dissatisfaction increases among female/non-binary students. The intended strengthening of overall satisfaction among female students is therefore not observed in the 2025 results.

\begin{figure}[!hbt]
    \centering
    \includegraphics[width=0.75\linewidth]{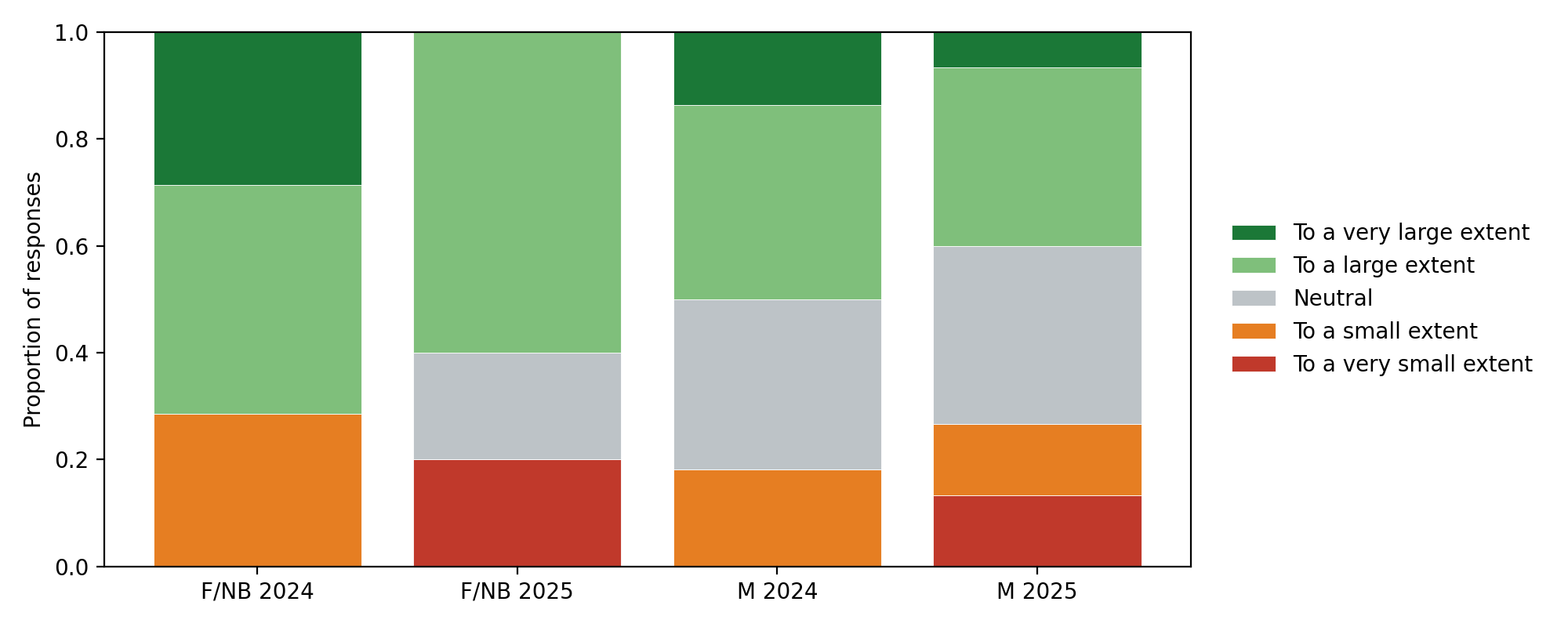}
    \caption{Provided with necessary prerequisites }
    \label{fig:req}
\end{figure}
\paragraph{Q8: To what extent did you feel were given the necessary prerequisites  to complete you assignments at the beginning of the semester?} This question assesses whether students felt prepared to complete assignments at the start of the semester, which is closely linked to confidence and early academic success. Compared to 2024, perceived preparedness decreases in 2025 for both genders, based on Figure \ref{fig:req}. A larger share of students report lower levels of preparedness, indicating that the overall perception of starting competence weakened rather than improved.

\begin{figure}[!hbt]
    \centering
    \includegraphics[width=0.75\linewidth]{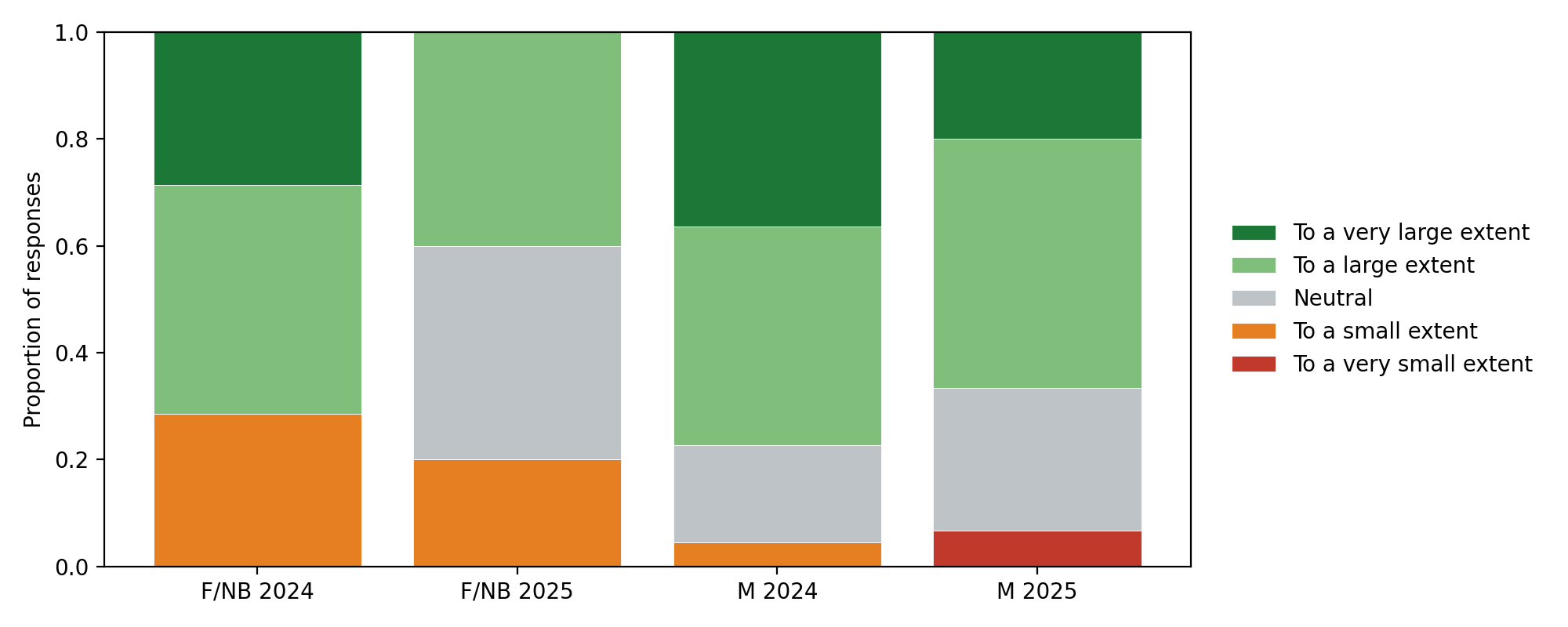}
    \caption{Comfortable asking questions}
    \label{fig:ask}
\end{figure}
\paragraph{Q9: To what extent do you feel comfortable asking academic questions to fellow students and teaching assistants?}
This question evaluates students’ comfort in seeking academic support, which is important for both inclusion and independent learning. Figure \ref{fig:ask} shows that, from 2024 to 2025, comfort levels decline for both genders. In 2024, female students already reported lower comfort than male students. In 2025, no female students report feeling comfortable to a very large extent. Overall, no improvement is observed in this area.

\begin{figure}[!hbt]
    \centering
    \includegraphics[width=0.75\linewidth]{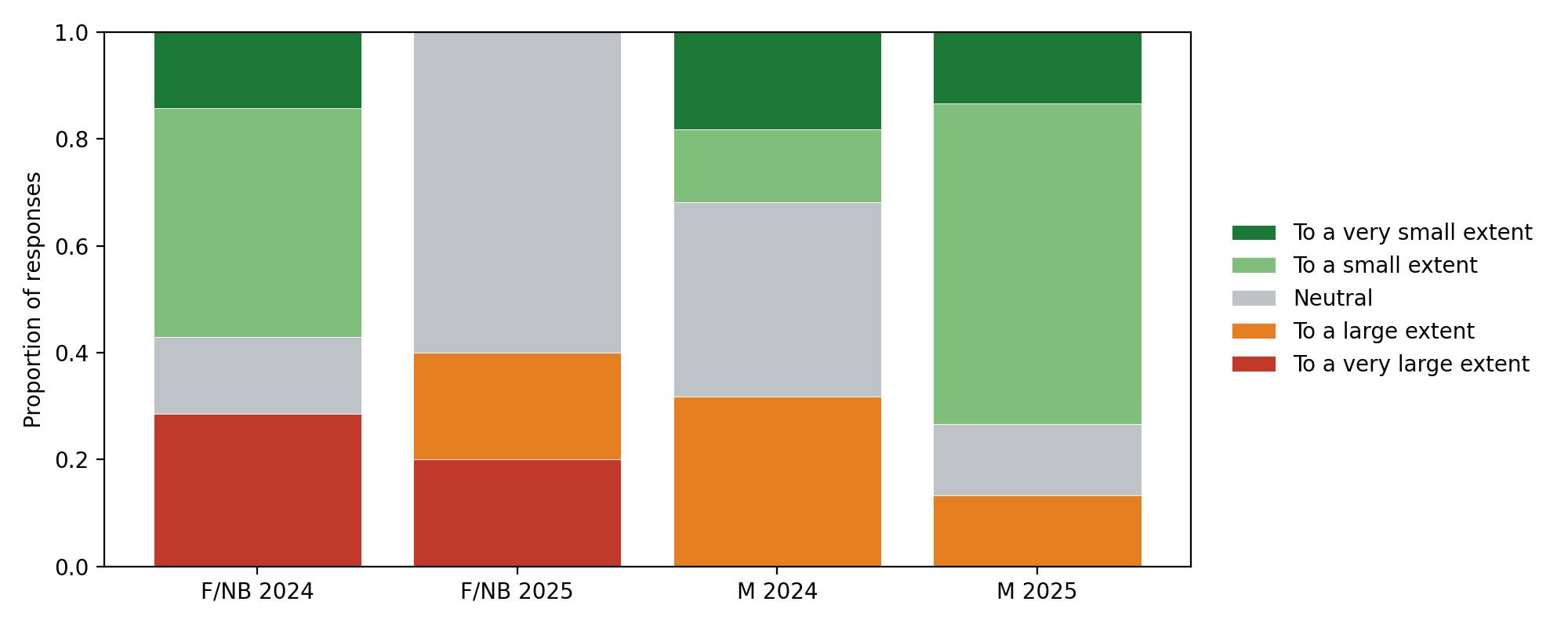}
    \caption{Feeling of being academically behind}
    \label{fig:behind}
\end{figure}

\paragraph{Q10: To what extent do you feel that you are academically behind your fellow students?}
This question measures students’ perception of academic comparison and confidence. Compared to 2024, the proportion of students who feel academically behind increases for female students in 2025, while it decreases for male students, according to Figure \ref{fig:behind}. In 2024, female students more often reported feeling behind than male students, and this gap increases in the following year. In 2025, this perception intensifies rather than decreases. The intended strengthening of academic confidence is therefore not reflected in this measure.

\begin{figure}[!hbt]
    \centering
    \includegraphics[width=0.75\linewidth]{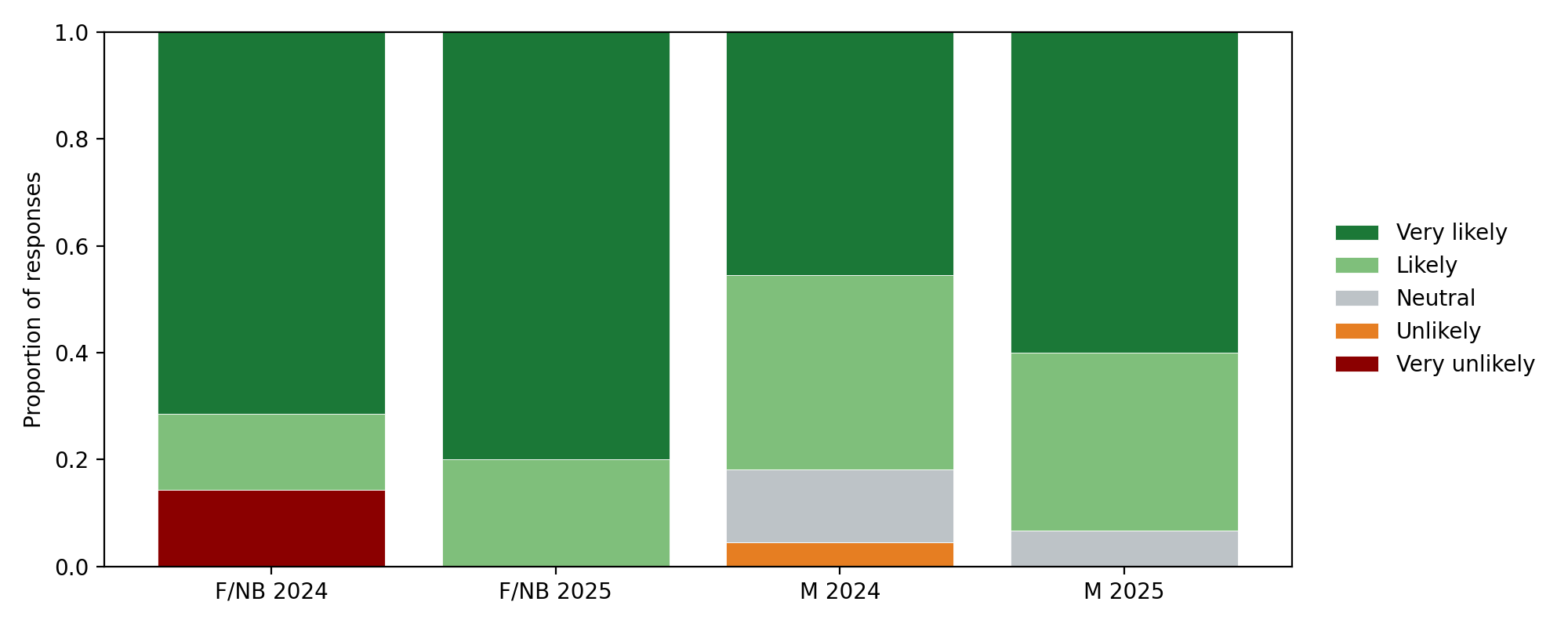}
    \caption{Likelihood to continue studying computer science}
    \label{fig:cont}
\end{figure}
\paragraph{Q11: How likely is it that you will continue studying computer science?}
This question captures students’ intention to remain in the program, which may indicate early dropout risk not visible in official statistics. Based on Figure \ref{fig:cont}, from 2024 to 2025, intention to continue strengthens for both genders. In 2024, more than 10\% of female students reported being unlikely or very unlikely to continue. In 2025, no female students report being unlikely to continue, and the majority report being very likely to continue. Male students also show increased certainty. This represents a clear positive development.

\begin{figure}[!hbt]
    \centering
    \includegraphics[width=0.75\linewidth]{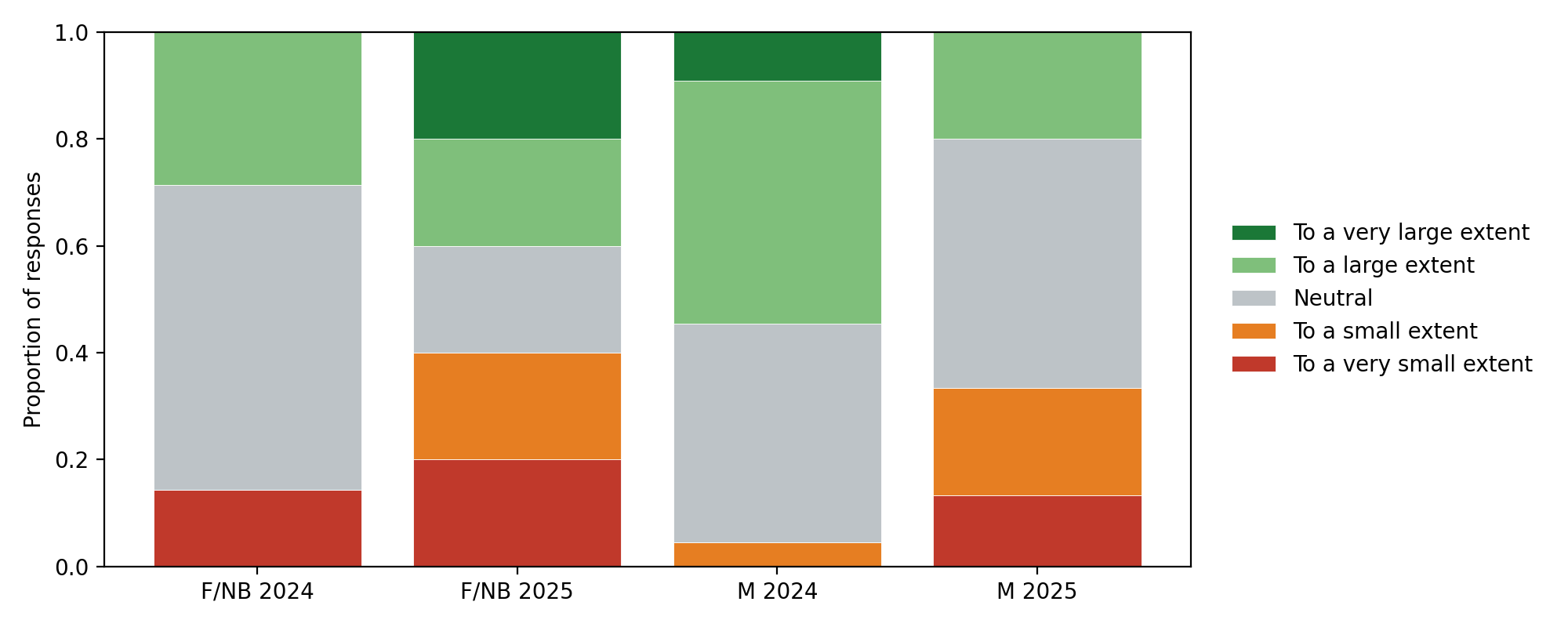}
    \caption{Introduced to inspiring topics and people within computer science}
    \label{insp}
\end{figure}
\paragraph{Q12: To what extent do you feel that you have been introduced to inspiring topics and people within the field of computer science through the CS department?}
This question measures perceived exposure to inspiring academic themes and individuals within the department. 
Figure \ref{insp} shows that, from 2024 to 2025, female students show a modest shift toward lower satisfaction, but with more responses in the higher categories. Male students show a slight shift toward more negative responses compared to 2024. Overall, exposure appears somewhat weakend among both groups.

\begin{figure}[!hbt]
    \centering
    \includegraphics[width=0.75\linewidth]{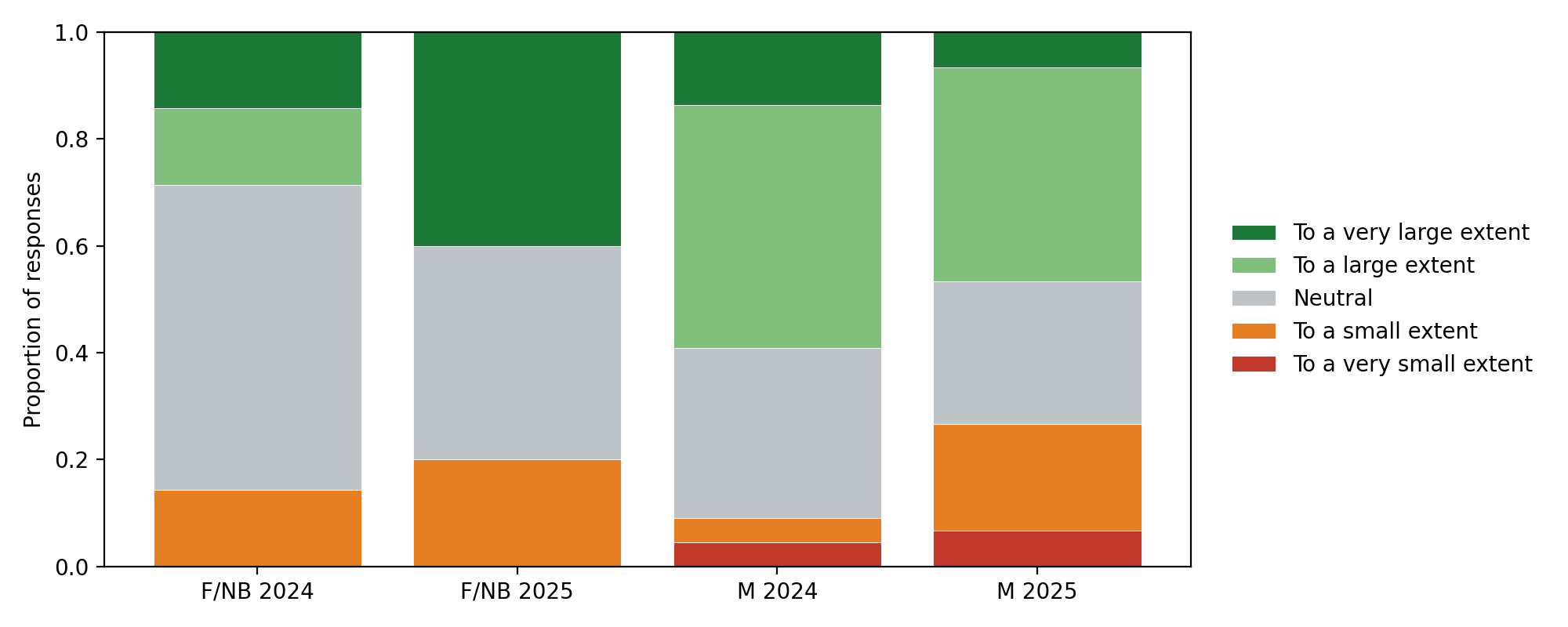}
    \caption{Exposure to diverse role models}
    \label{Rolemodels}
\end{figure}
\paragraph{Q13: To what extent do you feel that the CS department has exposed you to diverse role models?}
This question assesses students’ perception of diversity and representation within the department. Figure \ref{Rolemodels}
shows that, compared to 2024, female students show a clear shift away from neutral responses and toward higher categories in 2025. Male students also show a slightly negative shift. Overall, perceived exposure to diverse role models improves from 2024 to 2025 among female students, and the gender gap seems to be almost gone in 2025.

\subsubsection*{Conclusion about the impact of the cultural series}
The results from the cultural series present a more mixed picture compared to the results from the technical series. While exposure to inspiring topics and diverse role models shows modest improvement, particularly among female students, several other indicators do not improve from 2024 to 2025.

Perceived preparedness at the start of the semester decreases for both genders. Comfort in asking academic questions also declines, and feelings of being academically behind increase. Satisfaction with the study choice increases among male students but declines among female/non-binary students.

At the same time, one clear positive development is observed: intention to continue studying Computer Science strengthens for both genders, and no female students report being unlikely to continue in 2025.

%% file: sections/course_results.tex
Overall progression and dropout patterns provide important context for interpreting student outcomes across the two cohorts. The following tables summarize retention, exam participation, and performance by gender for 2024 and 2025.
\begin{table}[!hbt]
\centering
\begin{tabular}{lcc}
\hline
 & 2024 & 2025 \\
\hline
\textbf{Female} & & \\
Start (1st semester) & 11 & 6 \\
Start (2nd semester) & 10 & 5 \\
Dropout & 1 & 1 \\
\hline
\textbf{Male} & & \\
Start (1st semester) & 62 & 63 \\
Start (2nd semester) & 53 & 55 \\
Dropout & 9 & 8 \\
\hline
\textbf{Total dropout} & 10 & 9 \\
\hline
\end{tabular}
\caption{Dropout between first and second semester}
\label{dropout}
\end{table}

According to Table \ref{dropout}, the number of students who dropped out between the first and second semester is similar in the two cohorts, with ten students leaving in 2024 and nine in 2025. In both years, the majority of dropouts occur among male students, while only one female student leaves the program in each cohort. This suggests that the overall dropout pattern remains relatively stable across the two years.

\begin{table}[!hbt]
\centering
\begin{tabular}{lcc}
\hline
 & 2024 & 2025 \\
\hline
\textbf{Female} & & \\
Start & 11 & 6 \\
Coursework & 10 (90.9\%) & 6 (100\%) \\
Exam attendance & 9 (81.8\%) & 4 (66.7\%) \\
Passed & 9 (81.8\%) & 3 (50.0\%) \\
\hline
\textbf{Male} & & \\
Start & 62 & 63 \\
Coursework & 47 (75.8\%) & 50 (79.4\%) \\
Exam attendance & 44 (71.0\%) & 33 (52.4\%) \\
Passed & 36 (58.1\%) & 26 (41.3\%) \\
\hline
\end{tabular}
\caption{Progression from start of semester to passed exam}
\label{progress}
\end{table}

Most students who started the program completed the coursework requirements, particularly among female students, based on Table \ref{progress}. However, the proportion of students attending and passing the exam is lower in 2025 compared to 2024. The decline is most noticeable among male students, although female exam attendance and pass rates also decrease.

\begin{table}[!hbt]
\centering
\begin{tabular}{lcc}
\hline
 & 2024 & 2025 \\
\hline
\textbf{Female} & & \\
Passed & 9 & 3 \\
Failed & 0 & 1 \\
Attended & 9 & 4 \\
Pass rate & 100\% & 75.0\% \\
\hline
\textbf{Male} & & \\
Passed & 36 & 26 \\
Failed & 8 & 7 \\
Attended & 44 & 33 \\
Pass rate & 81.8\% & 78.8\% \\
\hline
\end{tabular}
\caption{Exam results among students who attended the exam}
\label{exam}
\end{table}

Table \ref{exam} shows that among students who attended the exam, pass rates remain relatively high in both cohorts. Female students show a decrease in pass rate from 100\% in 2024 to 75\% in 2025, although the number of female students is small. For male students, the pass rate remains relatively stable, with a slight decrease from 81.8\% to 78.8\%.

Both exams were graded as pass/fail; however, the exam formats differed between the two years. The 2024 exam was a home exam, where the use of large language models was not allowed, but not actively enforced. The 2025 exam was an on-campus exam, which likely better reflects individual performance under controlled conditions. Due to these differences in exam format and conditions, the results cannot be directly compared. The lower pass rate in 2025 therefore does not necessarily indicate reduced learning outcomes.

%% file: sections/analysis_of_results.tex
The evaluation of BBI 2.0 reveals a clear distinction between outcomes related to technical preparedness and outcomes related to confidence, comfort, and overall satisfaction. This distinction is consistent across the impact-survey results, participation patterns, and contextual exam data, and provides important insight into the strengths and limitations of the intervention. It is also important to interpret these findings in light of broader structural changes to the introductory course implemented between 2024 and 2025. These changes likely affected students’ overall experiences and may help explain the results more clearly.

The purpose of this section is to interpret the BBI initiatives in relation to the results presented, and to discuss what these findings reveal about the series impact.

\subsection{Effects of the Technical Series}

The most consistent positive developments are observed in areas directly addressed through the technical workshops. Students report improved perceived introductions to several foundational technical topics, including terminal usage, file systems, debugging, the code development cycle, and GitHub. In multiple cases, previously observed gender gaps are reduced among respondents in 2025.

These findings suggest that targeted, hands-on workshops can be effective in improving students’ perceived technical preparedness, particularly for students with limited prior experience. The greatest improvements are observed for workshops held earlier in the semester, which also had higher participation. This indicates that timing and early intervention are important factors in the effectiveness of technical support initiatives.

However, participation declined steadily over the semester, and workshops held later had substantially lower attendance. As a result, the improvements observed in later topics reflect the experiences of a smaller, more self-selected group of students. While the results are encouraging for participants, declining participation likely constrained the overall impact on the student cohort and limited the generalizability of these findings.

\subsection{Confidence, Comfort, and Sense of Belonging}

In contrast to the technical results, measures of academic confidence, comfort in asking questions, and overall satisfaction show limited improvement or even negative development. Female students in particular continue to report lower comfort levels and stronger feelings of being academically behind their peers, and in some cases, these perceptions worsen in 2025.

These results suggest that confidence and sense of belonging are shaped by broader structural and cultural factors within the study environment. This is consistent with prior research indicating that while technical skills can improve through short-term interventions, cultural factors such as sense of belonging and perceived inclusivity typically require sustained, long-term efforts to change \cite{Jaccheri2020, Frieze2019, Margolis2002}. Even when students report improved technical preparedness, this does not necessarily translate into increased confidence or comfort in academic interactions.

This patterns indicates that interventions targeting confidence and inclusion likely need to be implemented earlier, be longer-term, and be more closely integrated into the core structure of introductory courses.

One cultural event that appeared particularly valuable was the panel discussion involving both first-year and senior students. This format created an opportunity for dialogue between students at different stages of the program, allowing new students to ask questions and hear about the experiences of more advanced students. Events that facilitate interaction and discussion may therefore have greater potential to influence students’ sense of belonging than traditional presentation-based talks.

\subsection{Satisfaction and Retention Intentions}

Student satisfaction with the choice to study Computer Science shows mixed developments. While male students report increased satisfaction in 2025, female students report greater dissatisfaction than in 2024. This polarization suggests that early experiences affect students differently and that some students remain particularly vulnerable during the first semester.

At the same time, retention intentions improve in 2025. No female students report that they are unlikely to continue studying Computer Science, and the majority report that they are likely or very likely to continue. This suggests that dissatisfaction or low confidence does not necessarily lead to immediate dropout intentions.

One possible explanation is that increased awareness of career opportunities and exposure to diverse applications of Computer Science through the seminar series may encourage persistence even in the presence of challenges. Social factors, personal commitment, and external considerations may also influence students’ decisions to continue despite dissatisfaction.

\subsection{Exam Results and Structural Context}

Exam results provide important context for interpreting the impact-survey findings. The lower pass rate observed in 2025 compared to 2024 must be interpreted with caution, as the exam formats differed substantially between the two years. The 2024 exam was a home exam, while the 2025 exam was conducted on campus under controlled conditions.

Due to these differences, the exam results cannot be directly compared as indicators of learning outcomes. The lower pass rate in 2025, therefore, does not necessarily reflect reduced learning or weaker effects of the intervention. Instead, it highlights the importance of assessment design and exam conditions when interpreting performance data.

More broadly, as outlined in the introduction to this section, the structural changes to the introductory course between 2024 and 2025 likely influenced workload\footnote{The introductory course was restructured into two separate courses, with the first part enrolling a larger cohort of students from multiple study programs.}, expectations, and student experiences. These changes may help explain increases in dissatisfaction, reduced confidence, and lower exam pass rates in 2025, independently of the BBI 2.0 seminar series.

The exam results suggest that improved perceived preparedness and confidence in technical skills, as reported in some survey measures, do not automatically translate into improved exam performance in the short term. Structural factors, assessment format, and exam conditions likely play a significant role.

\subsection{Dropout Rate}
Reducing early dropout, particularly among female students, was one of the main objectives of the BBI 2.0 seminar series. The dropout rate is therefore an important indicator when evaluating the overall impact of the intervention. As presented in the results, the total number of students who dropped out between the first and second semester was somewhat the same for the 2024 and 2025 cohorts, especially among female students. The unchanged dropout rate indicates that the intervention alone was not sufficient to significantly influence early retention.

When comparing the results from the technical part of the seminar series with those from the cultural part, the findings suggest that increased understanding of certain technical topics is not necessarily sufficient to improve students’ overall confidence or sense of belonging. Since these factors are known to influence students’ decisions to continue or discontinue their studies, the absence of a reduction in dropout is not unexpected. These results therefore suggest that improving technical understanding alone may not be enough to address the broader factors that influence early dropout.

\subsection{The Number of Female Students}
Another important consideration is the lower number of female students starting the program in 2025 compared to 2024. This suggests that efforts to improve gender balance should not focus solely on retention within the first year, but also on increasing the number of female students entering the program. If fewer female students are recruited at the outset, improvements in retention alone will have limited impact on the overall gender balance. This indicates that recruitment and early outreach may be equally important areas to address in future initiatives.

Targeted efforts may also be more feasible given the smaller number of female students in the 2025 cohort. A reduced group size creates opportunities for more focused initiatives aimed specifically at supporting female students, as well as for closer follow-up with participants without requiring substantial additional resources. This aligns with the results from the technical workshops, which indicate that hands-on, targeted support can have a positive impact on students’ concrete knowledge and preparedness.

\subsection{Differences in Impact across Genders}
The results suggest that the series as a whole may have influenced female and male students’ experiences differently. In particular, female students appear to report increased exposure to inspiring topics and role models compared to the previous cohort, while the responses from male students remain more stable. This pattern indicates a modest improvement in how female students experience exposure to people and topics within computer science. 

It is important to note that the seminar series was open to all students. This decision was made in order to avoid gender exclusion and to ensure that the initiative contributed to a more inclusive learning environment. As a result, the findings and the discussion in this report are based on responses from both female and male students. 
The group of male students perceived the initiative as less representative of their own experiences, as the content and design of the series were primarily motivated by challenges previously identified among female students. While the intention was to address gender imbalance without excluding other groups, this perception highlights the complexity of designing interventions that aim to support underrepresented groups while remaining inclusive to the broader student population.

\subsection{Visibility of the Department}

In addition to its effects on students, the BBI 2.0 seminar series also appears to have contributed to increased visibility of the Department of Computer Science beyond the university. Through the cultural seminar component, a range of external speakers from academia and industry were invited to participate in the series. These interactions created opportunities to present the department, its study programs, and its learning environment to contributors from outside the institution.

While this was not a primary objective of the initiative, it represents a positive secondary outcome of the project. Establishing connections with external presenters may contribute to strengthening the department’s visibility at both the national and European level, and may support future collaboration, recruitment, and knowledge exchange. 

\subsection{Overall Assessment of Project Goals}

The results provide a mixed picture with respect to the goals of BBI 2.0. The most visible improvements are observed in students’ perceived preparedness for several foundational technical topics addressed through the workshops. These findings suggest that the technical component of the seminar series contributed positively to supporting students in the early stages of their studies, aligning with the goal of improving students’ perceived preparedness for introductory courses.

However, the results do not indicate similar improvements in measures related to academic confidence, comfort in asking questions, or overall satisfaction with the study choice. These outcomes contrast with the project’s stated aims of strengthening students’ confidence and increasing satisfaction with their choice of study, as outlined in the introduction. This suggests that such aspects of the study experience may be influenced by broader structural factors within the study program that extend beyond the scope of a seminar series.

While students’ stated intention to continue studying computer science strengthened between the cohorts, the overall dropout numbers remain similar, indicating limited impact on the primary goal of reducing early dropout.

%% file: sections/evaluation.tex
While the previous section evaluated the outcomes of BBI 2.0 in relation to the project’s goals, this section focuses on students’ experiences with the seminar series itself. The aim is not to assess whether the initiative achieved its intended effects, but to evaluate how the different components of the series were perceived by participants.

To support this, a separate evaluation survey was distributed to students who participated in the seminar series. The survey was conducted at the beginning of the Spring semester in 2026 
(i.e., the semester following the completion of the BBI 2.0 series) and included questions about both the technical workshops and the cultural events, as well as the overall experience. In total, seven students responded to the survey.

Due to the relatively small sample size, the results presented in this section should be interpreted as indicative rather than representative of the full student cohort. The findings are therefore used to provide qualitative insight into how the seminar series was experienced.

The plots in this chapter present the distribution of responses to each survey question. The horizontal axis shows the distribution of students in the different response categories, and the vertical axis shows the question.  

\subsection{The Technical Workshops}
\begin{figure}[!hbt]
    \centering
    \includegraphics[width=1\linewidth]{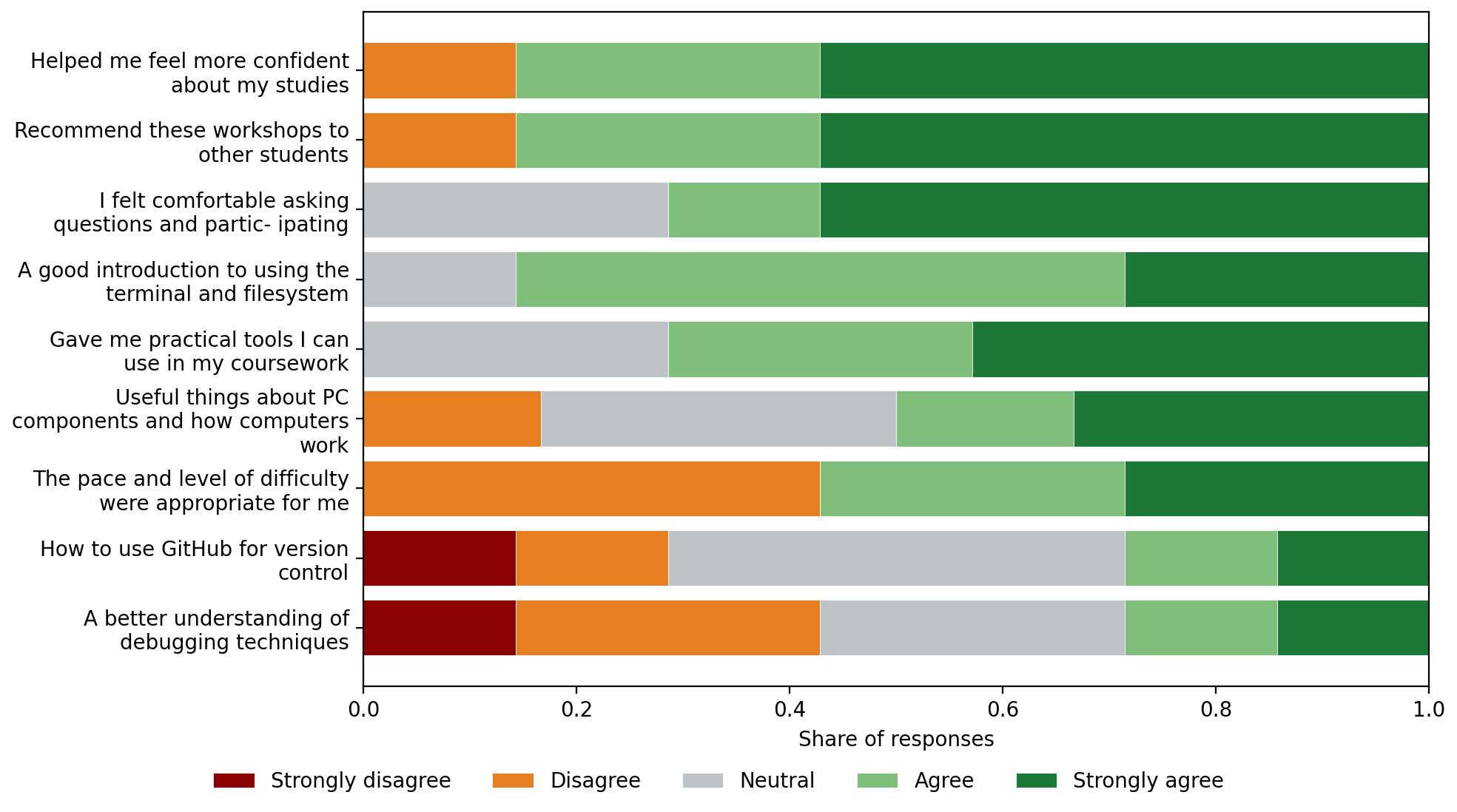}
    \caption{Evaluation of technical workshops}
    \label{fig:placeholder}
\end{figure}

Responses from the evaluation survey indicate that the technical workshops were generally perceived as useful and relevant. A majority of participants reported that the workshops helped them build confidence and that they would recommend them to other students.

The workshops on terminal usage and file systems were consistently highlighted as particularly valuable. These sessions appeared to provide practical knowledge that students found immediately applicable in their coursework.

At the same time, responses suggest variation in how the level and pace of the workshops were experienced. Some participants found certain topics, such as debugging, helpful for improving their understanding of code and problem-solving strategies. Others indicated that parts of the workshops covered material they already felt familiar with, while some topics, such as GitHub, were perceived as too brief to fully grasp.

Free-text responses further illustrate this variation. Several students emphasized that the workshops provided useful tools and helped structure their approach to programming tasks. Others suggested that more time could be allocated to advanced topics or practical exercises, particularly for tools used in collaborative work.

Overall, the evaluation survey suggests that the technical workshops were experienced as beneficial, particularly for developing foundational skills, while also reflecting differences in prior experience among participants.
    
\subsection{The Cultural Talks}
\begin{figure}[!hbt]
    \centering
    \includegraphics[width=1\linewidth]{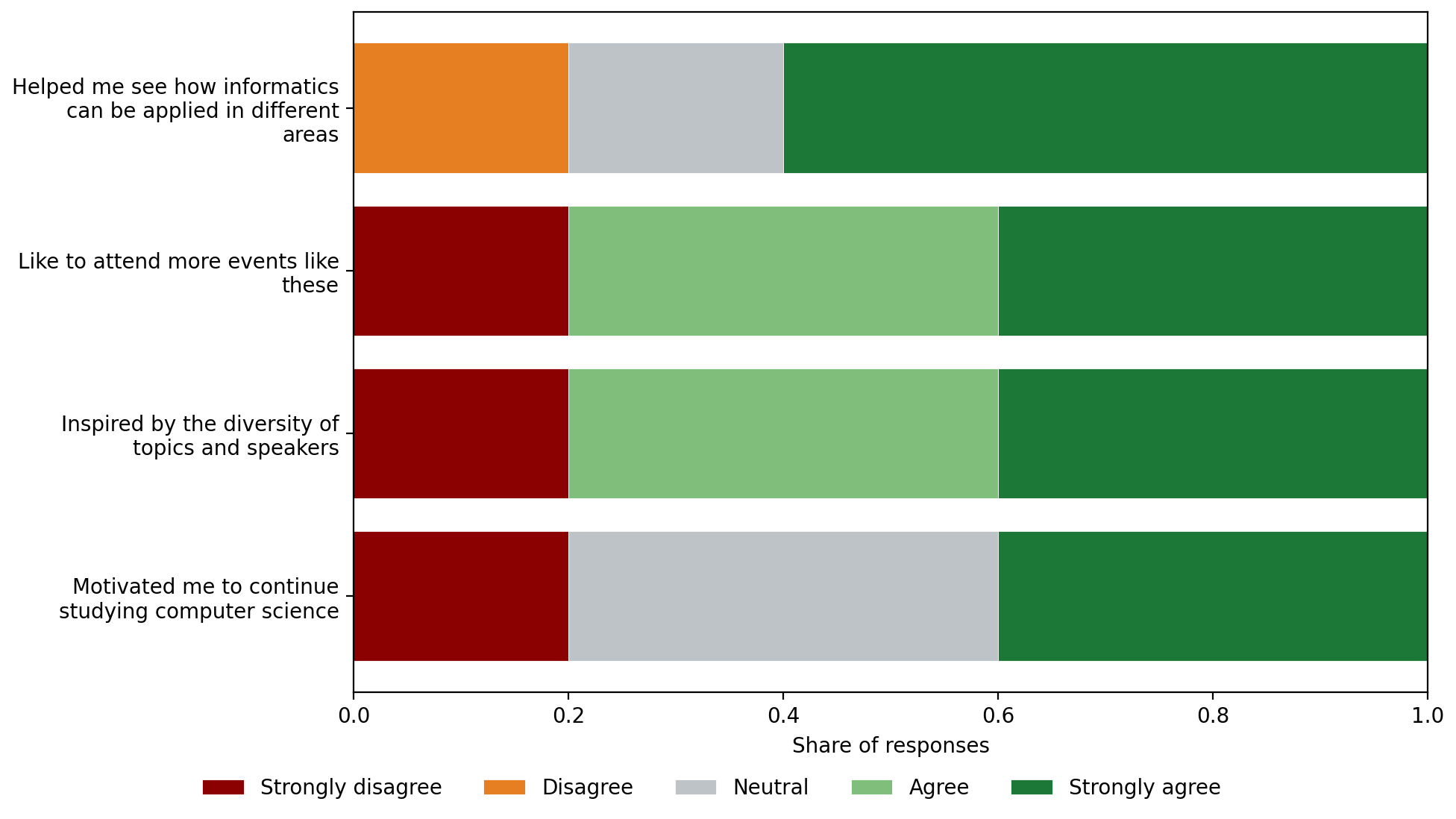}
    \caption{Evaluation of cultural talks}
    \label{fig:placeholder}
\end{figure}

Responses related to the cultural events indicate that participants valued being exposed to a broader view of Computer Science. Many students reported that the events helped them see how Computer Science can be applied in different domains and provided insight into possible career paths.

At the same time, responses related to motivation and identification with the field were more varied. Some participants reported that the events strengthened their interest in Computer Science, while others had more neutral experiences. Similarly, perceptions of representation among speakers were mixed, with some students feeling well represented, while others particularly male students felt less able to identify with the speakers.

Free-text responses reinforce the positive aspects of exposure and insight. Students noted that the talks helped them better understand the range of opportunities within Computer Science and appreciate the different ways computer science can be applied.

Overall, the cultural events were experienced as valuable for broadening perspectives, although their influence on motivation and identification with the field appears to vary across participants.

\subsection{Overall Evaluation of the BBI 2.0 Series}

\begin{figure}[!hbt]
    \centering
    \includegraphics[width=1\linewidth]{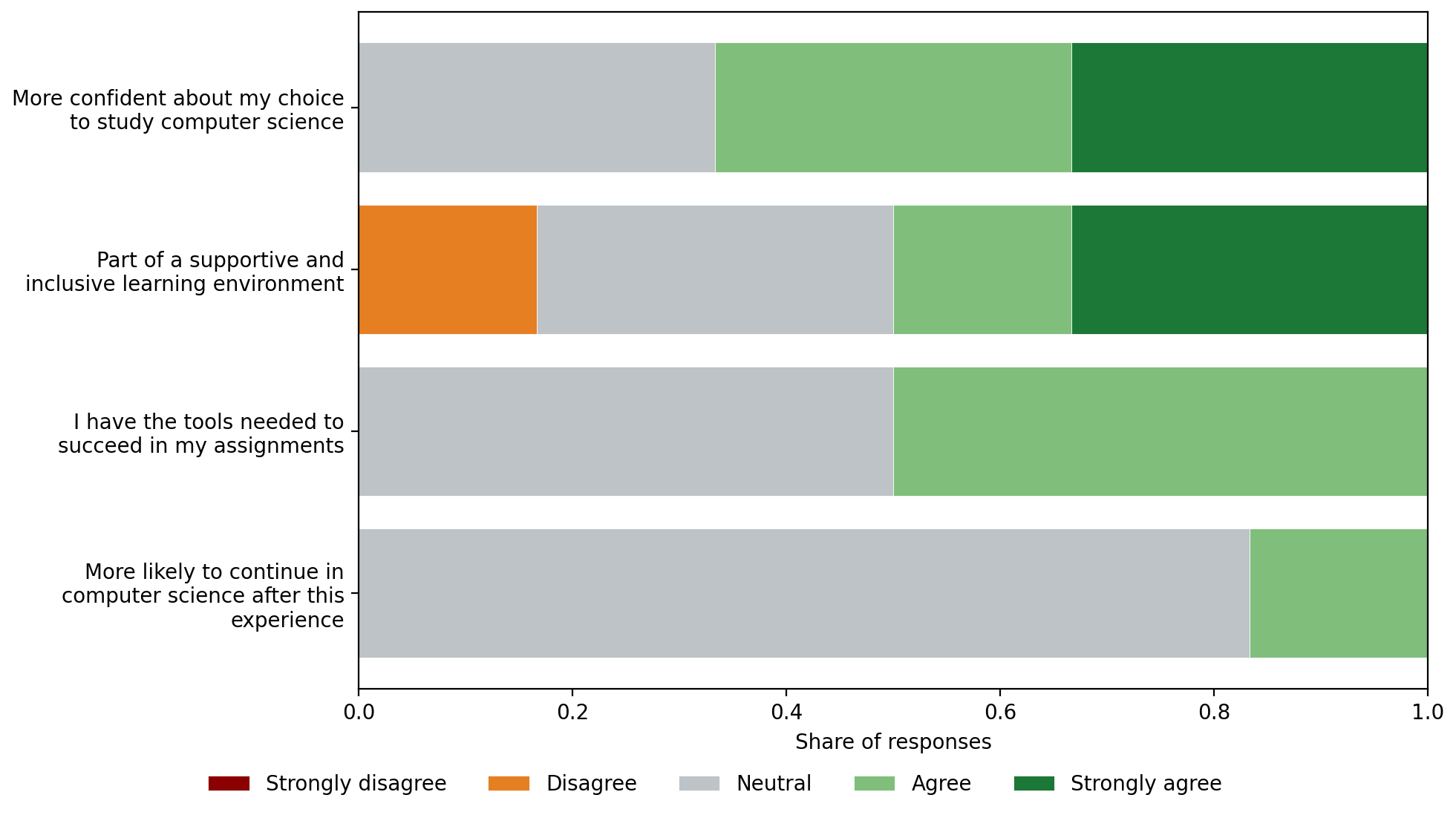}
    \caption{Overall evaluation}
    \label{fig:placeholder}
\end{figure}

The overall evaluation indicates that students generally perceived the seminar series as useful and relevant. Several participants reported increased confidence in their study choice and appreciated gaining practical tools and broader insight into the field.

At the same time, responses related to belonging, inclusiveness, and long-term motivation were more mixed. While some students experienced positive effects, others reported more neutral impressions.

Taken together, the evaluation survey suggests that the seminar series was most effective in providing practical tools and increasing awareness of the breadth of computer science, while having a more limited influence on broader aspects of the study experience.

%% file: sections/related_work.tex
Several intervention-based studies demonstrate that targeted measures can improve gender balance and retention in computer science education. A well-known example is the reform at Carnegie Mellon University, where beginner-friendly courses combined with broader structural changes increased female enrollment from 7\% to nearly 50\% without lowering academic standards \cite{Frieze2019}. Similarly, Happe et~al.~\cite{Happe2021} show that early, structured support can improve confidence and persistence, particularly for students with limited prior experience.

A central theme across the literature is the importance of inclusive teaching practices. Active, student-centered learning approaches increase engagement and foster a stronger sense of belonging among underrepresented groups \cite{Rainey2019}. Jaccheri et~al.~\cite{Jaccheri2020} further emphasize that improving gender balance requires not only pedagogical adjustments, but also structural and cultural changes. Through structured discussions, they identify student-informed interventions such as diversifying examples, improving classroom dynamics, and ensuring inclusive participation.

Mentorship, peer support, and exposure to role models are also consistently identified as key factors for retention. Supportive learning environments reduce feelings of isolation and improve persistence \cite{Margolis2002}, while exposure to role models can strengthen confidence and identification with the field \cite{Stout2016}. In addition, presenting computer science in a broader societal context has been shown to increase motivation and retention, particularly among women \cite{Margolis2002, Szlavi2021}.

At the same time, prior research highlights that different outcomes respond differently to interventions. While technical skills and confidence can improve through short-term efforts, cultural factors such as sense of belonging and perceived inclusivity are more deeply embedded and typically require sustained, long-term change \cite{Jaccheri2020, Frieze2019, Margolis2002}.

Taken together, these studies suggest that improving gender balance in computer science requires a combination of academic, social, and cultural interventions. Successful initiatives typically integrate early technical support, inclusive teaching practices, role models, and community-building activities—an approach that directly informed the design of the BBI 2.0 seminar series.

%% file: sections/conclusion.tex
This report set out to evaluate whether the BBI 2.0 initiative contributed to reducing early dropout, improving perceived preparedness, strengthening confidence, and increasing satisfaction with the study choice among first-year informatics students.

The findings show that these goals were only partially achieved and clearly point toward areas where further development is needed.

The most consistent positive outcomes are observed in students’ perceived technical preparedness. Improvements are reported across several foundational topics, and previously observed gender differences appear to be reduced. This indicates that the technical workshops successfully addressed gaps in prior experience and provided students with useful tools early in their studies.

In contrast, confidence, comfort, and satisfaction do not show similar improvement. Several indicators, including perceived preparedness at the start of the semester, comfort in asking academic questions, and feelings of being academically behind, remain unchanged or worsen in 2025. Satisfaction with the study choice does not improve among female students. These findings indicate that such interventions are not sufficient to influence broader aspects of the student experience.

The results also show no reduction in early dropout, although students’ stated intention to continue studying increases. This suggests that while the initiative may support motivation and persistence, its impact on actual retention remains limited within the timeframe of this study.

The fact that there are fewer female students in 2025 also needs to be considered. One possible approach is to not only focus on reducing dropout rates, but also to strengthen recruitment efforts. At the same time, the low number of female students may present an opportunity to organize smaller, targeted events aimed specifically at supporting female students.

Taken together, the evaluation shows that BBI 2.0 has been effective in addressing practical skill gaps, but less effective in influencing confidence, belonging, and retention. This distinction highlights an important implication: short, seminar-based interventions can improve technical preparedness, but broader challenges require more structural and sustained efforts.

Based on the findings presented in this report, the following recommendations are proposed:

\begin{itemize}
\item Activities should be scheduled as early as possible in the semester, as student engagement decreases over time.
\item Events that are closely aligned with ongoing coursework are more likely to attract participation and be perceived as relevant.
\item Improvements in confidence, belonging, and study satisfaction require longer-term efforts that extend beyond the timeframe of a short seminar series.
\item Addressing these broader aspects may require more structural and integrated changes within the study program, rather than relying on supplementary initiatives.
\item The panel debate highlighted the value of interactive and open formats. Events should therefore vary in format and include more dialogue-based sessions, as these appear more effective than traditional presentations in supporting engagement and belonging.
\item Greater focus should be placed on recruitment, in addition to organizing smaller, targeted events for female students.
\end{itemize}

Overall, the results indicate that future iterations of BBI should build on the strengths of the technical component, while expanding the initiative toward more integrated and long-term measures within the study program.

%% file: references.bib
@article{Frieze2019,
  author = {Frieze, Carol and Quesenberry, Jeria L.},
  title = {How Computer Science at CMU is Attracting and Retaining Women},
  journal = {Communications of the ACM},
  volume = {62},
  number = {2},
  pages = {23--26},
  year = {2019}
}

@article{Happe2021,
  author = {Happe, Lena and Buhnova, Barbora and Koziolek, Anne and Wagner, Ina},
  title = {Effective measures to foster girls’ interest in secondary computer science education},
  journal = {Education and Information Technologies},
  volume = {26},
  number = {3},
  year = {2021}
}

@article{Rainey2019,
  author = {Rainey, Kaitlin and Dancy, Melissa and Mickelson, Roslyn and Stearns, Elizabeth and Moller, Stephanie},
  title = {A descriptive study of race and gender differences in how instructional style influences STEM major choice},
  journal = {International Journal of STEM Education},
  volume = {6},
  number = {1},
  year = {2019}
}

@inproceedings{Jaccheri2020,
  author = {Jaccheri, Letizia and Pereira, Carla and Fast, Stig},
  title = {Gender Issues in Computer Science: Lessons Learnt and Reflections for the Future},
  booktitle = {2020 22nd International Symposium on Symbolic and Numeric Algorithms for Scientific Computing (SYNASC)},
  pages = {9--16},
  year = {2020}
}

@book{Margolis2002,
  author = {Margolis, Jane and Fisher, Allan},
  title = {Unlocking the Clubhouse: Women in Computing},
  publisher = {MIT Press},
  year = {2002}
}

@article{Stout2016,
  author = {Stout, Jane G. and Dasgupta, Nilanjana and Hunsinger, Megan and McManus, Melissa A.},
  title = {STEMing the Tide: Using Ingroup Experts to Inoculate Women’s Self-Concept in STEM},
  journal = {Journal of Personality and Social Psychology},
  volume = {100},
  number = {2},
  pages = {255--270},
  year = {2016}
}

@article{Szlavi2021,
  author = {Szlávi, Anikó and Bernát, Péter},
  title = {Young women’s barriers to choose IT and methods to overcome them},
  journal = {Teaching Mathematics and Computer Science},
  volume = {19},
  number = {1},
  pages = {77--101},
  year = {2021}
}
